\documentclass[useAMS,usegraphicx,usenatbib]{mn2e}

\usepackage{amssymb}
\usepackage{amsmath}

\newcommand{\be}{\begin{equation}}
\newcommand{\ee}{\end{equation}}

\newcommand{\kb}{$k_{b}$\,}
\newcommand{\mpr}{$m_{p}$\,}
\newcommand{\mtot}{$M_{cl}$\,}
\newcommand{\nj}{$\rho_{jet}$\,}

\newcommand{\ptr}{$p_{t}$\,}
\newcommand{\rhot}{$\tilde{\rho}$\,}

\newcommand{\rtr}{$r_{t}$\,}

\newcommand{\nel}{${\rm n}_{e}$\,}

\title[AGN Jet-induced Feedback in Galaxies]{AGN Jet-induced Feedback in Galaxies. I. Suppression of Star Formation.}
\author[V. Antonuccio-Delogu, J. Silk]{V. Antonuccio-Delogu$^{1,2}$\thanks{van@astro.ox.ac.uk},
  J. Silk$^{1}$\\
$^{1}$Astrophysics, Department of Physics, University of Oxford, Keble Road Ox1 3RH, Oxford, United Kingdom\\
$^{2}$INAF - Osservatorio Astrofisico di Catania, Via S. Sofia 78, Catania, I-95123, Italy}

\begin{document} 

\date{Accepted ??. Received ??; in original form 2007 ??}

\pagerange{\pageref{firstpage}--\pageref{lastpage}} \pubyear{2007}

\maketitle

\label{firstpage}

\begin{abstract}
Relativistic jets originating from Supermassive Black Holes
(SBHs) can have a considerable impact on the Interstellar/Intergalactic Medium
(ISM/IGM) within which they
propagate. Here we study the interaction which a relativistic jet,
and the cocoon associated with its penetration into the ISM,
has on the evolution of a dense cloud, placed very near the cocoon's path, by
analyzing a series of high-resolution
numerical simulations, and studying the dependence 
on jet input power, between $P_{jet} =
10^{41}-10^{47}\, {\rm erg/sec}$.  The density Probability Distribution
Function (PDF) within the cocoon can be described in terms of
two distinct components, which are also spatially distinct: a 
low- and a high-density component. The former is associated with the
shocked gas within the internal region of the cocoon, while the
latter is associated with the outer, shocked region of the cocoon
itself. The PDF of the post-shocked region is well approximated by a
modified 
lognormal distribution, for all values of $P_{jet}$. During the \textit{active} phase, when the
jet is fed by the AGN, the cloud is subject both to compression and
stripping, which tend to increase its density and diminish its total mass. When the jet is switched off (i.e. during the \textit{passive}
phase) the shocked cloud cools further and tends to become more
filamentary, under the action of a back-flow which develops within the
cocoon.\\
We study the evolution of the star formation rate within the cloud,
assuming this is determined by a Schmidt-Kennicutt law, and we analyze the different physical
factors which have an impact on the star formation rate. We show that,
although the star formation rate can occasionally increase, on time
scales  of the order of $10^{5}-10^{6}$ yrs, the star formation rate
will be inhibited and the cloud fragments. The cooling time of the
environment within which the cloud is embedded is however very
long: thus, star formation from the fragmented cloud remains strongly
inhibited. 

\end{abstract}


\section{Introduction}

One of the most intriguing research areas in contemporary extragalactic
astrophysics involves the study of the interplay between nuclear Black
Holes (BHs) in galaxies, the (relativistic) jets which they can
produce, and the Interstellar/Intergalactic Medium (ISM/IGM) within
which they propagate. Observation and modeling of the
propagation of jets within the ISM is an important part of this effort.\\ 
Since the seminal works of \citet{1974MNRAS.166..513S} and
\citet{1991MNRAS.250..581F}, much effort has been dedicated to the
study of the propagation of a jet into the Interstellar Medium
(hereafter ISM), and to its consequences for the detectability of the
jet. Relatively less attention has been paid to the impact of the
jet and the cocoon generated by it on an inhomogeneous ISM. 
\citet{2005MNRAS.359..781S}
have performed numerical simulations of the interaction of a cocoon with
a set of clouds embedded within a diffuse ISM, mostly paying attention to the evolution of the
jet's morphology. More recently,
\citet[][hereafter KA07]{2007MNRAS.376..465K} have also studied the turbulence
induced by the interaction of jets with cold clouds embedded in an
Interstellar/Intergalactic medium. In their simulations, they show
that the jet is Kelvin-Helmholtz unstable, and can shear a cold
cloud embedded in the ISM/IGM, thus inducing turbulence. Their
simulation setup is however different from ours, mainly because they
focus their attention on a small portion of the jet to study in detail
the interaction with IGM clouds and the resulting turbulence. However, as
we see in our simulations, turbulence in the cocoon arises naturally due
to the general dynamical expansion of the cocoon itself. \citet{2007ApJS..173...37S,2007Ap&SS.311..293S} have
taken into account the presence of large-scale density gradients in
the ISM distribution, and the ability of a jet to emerge out of a
galactic disc. They notice that the jet can eventually percolate
through the inhomogeneous ISM, and emerges with a very disturbed morphology.\\
All the papers quoted made use of fixed-mesh numerical codes:
in this paper, we use an Adaptive Mesh Refinement (AMR) code to
follow in detail the evolution of turbulence within the cocoon
produced by the jet during its propagation within the
ISM/IGM. Allowing for a high
refinement level, we can resolve small turbulent eddies, and study
their statistical properties, and how they affect the evolution of a
cloud embedded within the cocoon. We focus our attention on star
formation within the cloud, and on the evolution of the density field
within the cocoon, as our main aim is to understand how,
within the turbulent cocoon produced by typical AGN jets at moderately
high redshifts (z$\approx 1-2$), this turbulence can modify the
ongoing star formation within the cloud.\\
The paper is organized as follows. In section~\ref{comp_fram} we briefly describe the numerical and
computational set-up. In section~\ref{comp_fram}, we present the
numerical techniques and physical ingredients to simulate the jet and
the ISM. We
continue in section~\ref{in_conf} by presenting the initial
configuration and the physical parameters space spanned by the
simulation, and we also discuss issues of  numerical resolution. In
section~\ref{ev_cocoon} we describe the results of the simulations and the
evolution of the cocoon. We consider the evolution during both the \textit{active}
phase, while the jet is present, and the subsequent \textit{passive}
phase. Then, in section~\ref{ev_pdf}, we
discuss the density probability distribution function (PDF) of the
matter within the cocoon, and we
find that it is well described by a modified lognormal function. In
section~\ref{jet_cloud} we then study the evolution of the Star
Formation rate (SFR) within the cloud. In section~\ref{sec_disc} we
discuss the implications of these simulations, comparing our results
with those of similar papers. Conclusions are presented in the
final section.\\
In the following, we adopt cgs units, unless otherwise 
explicitly stated, but we adopt kpc for lengths. The underlying
cosmological parameters are taken from the 3-year WMAP best fit
$\Lambda$CDM model \citep{2006NewAR..50..850B}: ${\rm H}_{0} =
74^{+3}_{-3}$ Km/sec/Mpc, $\Omega_{m} = 0.234\pm 0.035$,
$\Omega_{b}h^{2} = 0.0223\pm 0.0008$. The unit of time is taken to be
the Hubble time for this model, i.e.: $t_{0} = 1.35\times 10^{10}$ years. $G$, \kb and \mpr denote the
gravitational constant, the Boltzmann constant and the proton mass,
respectively.\footnote{
For the interested reader, we have put some animations of these simulations on the website: http://web.ct.astro.it/cosmoct/web\_group/research.html.}\\

\section[]{Computational framework}\label{comp_fram}
To perform the simulations described here, we have used FLASH v. 2.5 
\citep{2000ApJS..131..273F}, a parallel, Adaptive Mesh Refinement code which
implements a second order, shock-capturing PPM solver. FLASH's modular
structure allows the inclusion of physical effects like external heating,
radiative cooling and thermal conduction (among others). The user is
permitted considerable freedom in the specification of the
refinement criteria, which
can be customized to reach very high spatial and temporal
resolutions in selected regions.\\
\noindent
The main purpose of this work is to
study in detail the interaction of a relativistic jet and the cocoon
which it generates with pre-existing clouds, and how it can affect star
formation within the cloud itself. Ideally, a full 3D simulation
should have a spatial resolution high enough to resolve {\em both} the
turbulent motions within the cocoon and the thermodynamic structure of
the cloud, until the end of the simulation. The former task is
important when one realizes that, in addition to the direct
interaction with the jet, the cloud is also significantly affected by
the random interactions with the turbulent eddies present within the
cocoon. The computational requirements imposed by this task, however,
are prohibitive, when one considers that the smallest turbulent cells to be
resolved should have a size comparable to that of the cloud (10
$h^{-1} pc$ here), and the computational box is between 2 and 4 $\times 10^{2}$ times
larger. For this reason, we have restricted ourselves to 2D
simulations, where this resolution can easily be reached.\\
\noindent
We have included radiative
cooling, described by a standard cooling function with  half solar
metallicity \citep{1993ApJS...88..253S}, extended to high temperatures
(T $> 10^{7}$ K, see appendix~\ref{append_2}). Gravity is also included, while
thermal conduction and magnetic fields are not. The former could
possibly be relevant for the evolution of the cocoon, although on
timescales longer than those considered here
\citep{2007MNRAS.376..465K}. Regarding magnetic fields, the evidence
is that, if present within the diffuse IGM, their magnitude is not
larger than a few microgauss, thus making the IGM a high-$\beta$
plasma, and the magnetic field would then not significantly affect the global dynamical
evolution.\\
In the simulations the jet is modeled as a one-component fluid, with
a density  \nj
which is a fixed ratio $\epsilon_{j}$ of the environmental density
$\rho_{env}$. In order to suppress the growth of
numerical instabilities at the jet/IGM injection interface, we adopt
a steep, but continuous and differentiable transverse velocity and density
profile previously adopted in simulations of jet propagation \citep{1994A&A...283..655B,2004A&A...427..415P, 2005A&A...443..863P}:
\be
v_{x,j} = \frac{V_{j}}{\cosh\left\{\left(y-y_{j}\right)^{\alpha_{j}}\right\}}
\label{eq:vinjet}
\ee
\be
n_{j} = n_{env} - 
 \frac{\left(n_{env} - n_{j}\right)}{\cosh\left\{\left(y-y_{j}\right)^{\alpha_{j}}\right\}}
\label{eq:rhojet}
\ee
where: $\alpha_{j}=10$ is an exponent which determines the steepness of the
injection profile and $n_{j}, n_{env}$ denote the jet and environment number densities. 
This initial profile is highly sheared, and peaks
around $y_{j}$, with most of the thrust $n_{jet}v_{jet}^{2}$
concentrated around the center of the profile.   The presence of a
highly  sheared injection profile forces the code to refine the grid
structure, populating the injection region with subgrid meshes, thus
preventing the formation of numerical contact instabilities. In all
our runs, the injection point of the jet is chosen to lie at
the midpoint of the left boundary.
\section[]{Simulations}\label{in_conf}
Our simulations are characterized by seven parameters.
Two of these characterize the diffuse ISM: $n_{env},\, T_{env}$, two
more the cloud: $r_{cl},  M_{cl}$, and finally three characterize the
jet: $n_{j},\, y_{j},\, V_{j}$. As the SFR is mostly determined by the
strength of the shock within the cloud, as
previous models seem to suggest \citep{1994ApJ...420..213K}, we have
decided to mostly vary
the jet's input power $P_{jet} = 0.5 A_{j}m_{H}n_{j}V_{j}^{3}$ which,
together with the density contrast, determines the timescale of the cloud's
disruption. We keep most of the other parameters fixed: ISM density and temperature
at 1 e$^{-} {\rm cm}^{-3}$ and $10^{7}$ K, respectively, typical of
the hot ISM in the central parts of a massive elliptical at high
redshift ($z\approx
1$). Only in one run we have increased the size of the box, in order to
check that the main features of the evolution do not depend on
boundary conditions. All runs except one have been performed on a
relatively small box (4 $h^{-1}$ kpc), while in
run H2 we have used a box twice as large.\\
\noindent
In Table 1 we summarize the
main parameters of the different runs.  The
ratio between jet and environment density, {\rm n$_{j}$/{\rm
    n}$_{env}$} is fixed to $2\times 10^{-2}$ in runs using small
boxes, and decreased to a slightly lower value ($10^{-2}$) in run H2. The injection
region has a width of 100 $h^{-1}$ pc in runs in the small box, and
twice as much in run H2. The parameter $V_{j}$ in eq.~\ref{eq:vinjet},
is computed once $\rho_{j}, d_{j}$ and $P_{jet}$ are assigned. Finally,
we have chosen {\em open} boundary conditions, so the gas is free to flow
out of the simulation box. The implication of this is that gas is not
allowed to re-enter the box, so circulation motions on scales larger
than the simulation box cannot be reproduced.
\\
\noindent

\begin{figure}
\centering
\includegraphics[scale=0.35,angle=0]{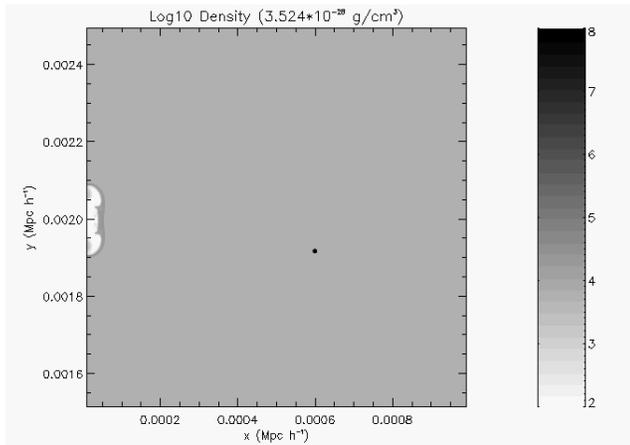}
\caption{Initial configuration for the runs. The plot is a magnification of the
  actual simulation box, showing a small region of the larger
  simulation box. The small, dense cloud lying along
  the path of the jet, placed at (0.6, 1.92) (in $\, {\rm kpc}\, h^{-1}$).}
\label{fig1}
\end{figure}
\subsection[]{Initial configuration}\label{sec:in_conf}
We place a small, dense cloud at a position
slightly offset from the jet's propagation direction (see
Fig.~\ref{fig1} and Table~\ref{tab:cl}). In order to keep the same
spatial resolution, the radius of the cloud is doubled in run H2,
where the simulation box is larger: consequently also the initial mass
of the cloud is larger.
\begin{table*}
 \centering
 \begin{minipage}{100mm}
  \caption{Parameters of the simulation runs. Columns from left to
    right are as follows: run label, simulation box size, background
    density (runs H1-3) or central density (runs NFW),
    jet/background density ratio, jet power. }
  \begin{tabular}{lcccc}
  \hline
   run & {\rm L}$_{box}$ (kpc$\, h^{-1}$) & n$_{env}$ (cm$^{-3}$) & n$_{jet}$/n$_{env}$ & {\rm W}$_{jet}$ (\rm{erg/sec}) \\
 \hline
 H0 & 4 & 1 & $0.02$ & $4\times 10^{40}$ \\
 H1 & 4 & 1 & $0.02$ & $8.61\times10^{41}$ \\
 H2 & 8 & 1 & $0.01$ & $1.34\times10^{44}$ \\
 H3 & 4 & 1 & $0.02$ & $10^{45}$ \\
 H4 & 4 & 1 & $0.02$ & $10^{46}$ \\
\hline
\end{tabular}
\end{minipage}
\label{tab:cl}
\end{table*}
We have chosen to study the effect of the jet
on a cloud located very near to its path, because this allows us to
check to what extent star formation is affected by the jet under the most
extreme conditions. In a forthcoming paper we will study a more
realistic setup, where we distribute a set of clouds with a realistic
mass spectrum, and study how star formation changes according to
the relative position within the cocoon associated with the jet.\\
\begin{table*}
 \centering
 \begin{minipage}{100mm}
  \caption{Cloud parameters. All distances are expressed in ${\rm kpc}
   \, h^{-1}$. From left to
    right, columns are as follows: simulation box size, x and y
   coordinates, cloud's radius and mass, the latter in ${\rm M}_{\sun}$.}
  \begin{tabular}{ccccc}
  \hline
   {\rm L}$_{box}$ (kpc$\, h^{-1}$) & x & y & ${\rm r}_{cl}$ & ${\rm M}_{cl}$ \\
 \hline
  4 & 0.6 & 1.92 & 10 & $1.65\times 10^{8}$\\
  8 & 1.2 & 3.84 & 20 & $1.30\times 10^{9}$\\
\hline
\end{tabular}
\end{minipage}
\end{table*}
\noindent
Most of the runs were evolved up to $\approx 10^{7}\,$ yrs.,while the
jet was active and supplying energy to the cocoon. Only in one run
(H4) the jet was switched off after $10^{7}\,$ yrs., and the further
evolution of the system was followed until the cloud was completely
destroyed.

\subsubsection[]{Model of embedded cloud}
We have chosen a model for the 
structure of the embedded clouds suggested by  
the numerical simulations performed by
\citet{2005ApJ...630..689B}, because the physical ingredients of these
simulations are likely to be representative of the
physical processes present in the ISM/IGM,  These simulations have been 
devised to provide a reasonable model for clouds
embedded within the IGM. Different sources of heating (UV-background,
the wind and radiative flux from the central QSO itself) provide
a significant energy input which can promote the formation of
pressure-confined clouds through thermal
instability. \citet{2005ApJ...630..689B} have shown 
that, for typical IGM density and temperatures, similar to those considered
in the present paper, the cooling time is a small fraction of the
dynamical time, and the ISM is prone to the development of small,
pressure-confined clouds. 
These are almost isothermal, with temperatures near the
lower extreme of the cooling function ($T_{cl}\approx 10^{4}$ K). 
In the mass range $10^{2.5}\leq 
M_{cl} \leq 10^{7} M_{\sun}$ they find a relationship between \rtr and the total mass
$M_{cl}$:
\be
r_{t} = \lambda M_{cl,4}^{\beta} \label{eq:cleq:2}
\ee
where we have defined: $ M_{cl,4} = M_{cl}/10^{4} {\rm M}_{\sun}$. If $r_{t}$ is measured in pc, we obtain: $\lambda = 28.87, \beta =
0.28\pm 0.04$. The upper limit of this mass range corresponds to the
Jeans mass for these clouds, implying that they are \textit{not self-gravitating}.\\
As we show in Appendix~\ref{append_1}, a reasonable model for these clouds is 
the \textit{Truncated Isothermal Sphere} \citep[TIS][]{1999MNRAS.307..203S}. 
Given the cloud's mass \mtot, the final
parameters of the configuration will depend on the background ISM 
thermodynamic state, and we assume that the latter is described by an
unperturbed ideal equation of state with density and temperature
$\rho_{cl}$ and $T_{cl}$, respectively, as
appropriate to a high temperature, low density, fully ionized 
plasma \citep[e.g.][]{1987smh..book.....P}. 
We denote the solution of the TIS equation by $F(\zeta)$, so that the cloud density
can be written as: $\rho_{cl} = \rho_{0}F(\zeta)$, $\rho_{0}$ being the
normalization factor and $\zeta = r/r_{0}$ a normalized
distance. The cloud's structure is entirely specified by assigning $r_{0},
\rho_{0}$ and the truncation distance $r_{t}$. Once the latter is
determined by the radius-mass relation found by
\citet{1999MNRAS.307..203S}, the two remaining factors are determined
by imposing  pressure equilibrium with the IGM, as shown in
Appendix~\ref{append_1}.
\section[]{Star formation}
Although our maximum spatial resolution could allow us to resolve
subparsec scales, we cannot follow star formation in any detail,
because this would require the inclusion of more physics (for instance
a very detailed treatment of molecular cooling) and temporal
resolution a few orders of magnitude higher than the one we have
adopted. Instead, we assume that the cloud is converting gas into
stars at a rate
 determined by the Schmidt-Kennicutt law
 \citep{1963ApJ...137..758S,1959ApJ...129..243S,1998ApJ...498..541K}:
 $\dot{\Sigma} = A\Sigma^{n}$, with: $A = (2.5 \pm 0.17)\times 10^{-4}\,
 {\rm M}_{\sun}\, yr^{-1} \,{\rm kpc}^{-2}\,$, $n = 1.4 \pm 0.15$. At any time, we
 assume that star formation proceeds only within
 those regions of the simulation volume where the following criteria
 are satisfied:
\begin{enumerate}
\item  Mass is larger than the Bonnor-Ebert mass;
\item Temperature is less than a prescribed upper cutoff, i.e. we
  assume that star formation is sharply
  inhibited in regions having $T \geq T_{c} = 1.2\times 10^{4} {\rm
  K}$.
\end{enumerate}
The Bonnor-Ebert mass \citep{1955ZA.....37..217E,1956MNRAS.116..351B}
is defined as the largest mass which a
pressure confined, gravitating cloud can reach before becoming unstable:
\be
M_{be} = 1.18 \frac{c_{s}^{4}}{\sqrt{G^{3}p_{ext}}} = 23.55
\frac{T_{4}^{2}}{p_{ext}^{1/2}} \, M_{\sun}\label{eq:sf:1}
\ee
where $p_{ext}$ is the pressure at the surface of the cloud,
temperature is expressed in units of $10^{4}$ K, and we
have assumed that an isothermal equation of state applies, so that
the sound speed is given by: $c_{s} = (k_{B}T/m_{p})^{1/2}$. The
temperature cutoff is often
adopted in star formation models to exclude regions where the UV flux
would be too high to permit significant star formation. Our
criteria to select the star forming region are very similar to
those adopted by \citet{2008MNRAS.383.1210S}.\\
From now on, all characteristic properties of the cloud such as its mass
or size will always be referred to the {\em star forming
  region} as just defined.

\subsection{Numerical resolution}
As our main goal is that of studying in detail the evolution of star
formation within the cloud, we need sufficiently high spatial
resolution during each run. The outer parts of the cloud
are stripped due to interaction with the jet, and it is in principle
difficult to predict how much the cloud's size and mass will be
reduced during the evolution. For this
reason, we have applied a refinement criterion which will enable us to
get high resolution even during the late phases, when the star forming
region is considerably reduced.
\begin{figure}
\centering
\includegraphics[scale=1.0,angle=0]{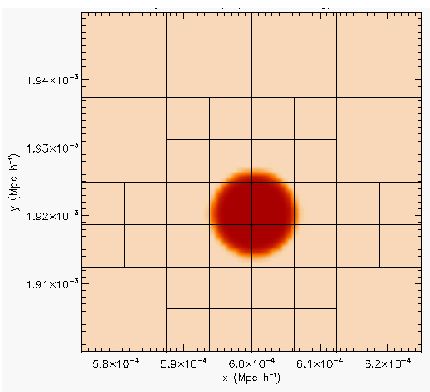}
\includegraphics[scale=1.0,angle=0]{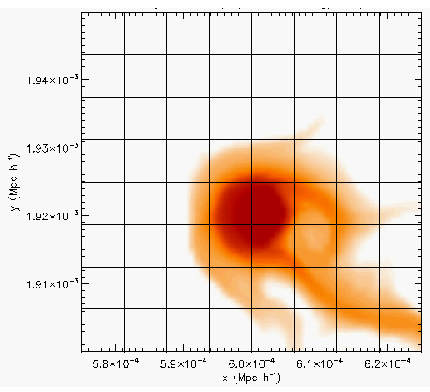}
\caption{Computational block distribution for two different steps of
  run H1, taken at $t = 2.6\times 10^{5},\, upper$ and $5.3248\times
  10^{5}, \,lower$ (time in yrs).  Each little square represents a FLASH {\em
  block}, and each of
  these contains 64 mesh cells (not reproduced in the figures). Note
  that the size of the region is 50 $h^{-1}$ pc, a small
  fraction of $L_{box}$, and the blocks represented are at the highest
  refinement level. }
\label{fig1res}
\end{figure}
The spatial resolution in FLASH is determined by three parameters: the
number of blocks in the initial decomposition, which in 2D simulations
is given by: $N_{blx}\times N_{bly}$, the number of mesh cells
within each block, $nb_{x}\times nb_{y}$, and the maximum
allowed refinement level $l_{r}$. In the present work we have chosen:
$N_{blx}=N_{bly}=20$, $n_{bx}=n_{by}=8$, and $l_{r}=6$. Thus, the
smallest spatially resolved scale (down to mesh level) along each direction is
given by: $\Delta_{x} = \Delta_{y} = \left({\rm
  L}_{box}/N_{blx}\right)/\left(2^{(l_{r}-1)}n_{bx}\right)$, which,
for ${\rm L}_{box}=4 h^{-1}$ kpc gives: $\Delta_{x} = 0.78125 \,
h^{-1}{\rm pc}$. Note that the maximum refinement level around the
cloud is reached already at the start of the simulation, because of
the refinement criterion adopted. The approximate number of mesh cells contained within a cloud of radius
$R_{cl}$,  $n_{m}$, can be estimated as: $n_{m} = int\left(\pi
R_{cl}{2}/{\Delta A_{m}}\right)$, where $\Delta A_{m} =
\Delta_{x}\Delta_{y}$ is the area of a single mesh. In our case
initially $R_{cl} \simeq 10 h^{-1}$ pc, so we get: $n_{m} \approx
514$. A typical block distribution is shown in Figure~\ref{fig1res}.\\ 
By default, FLASH refines the grid at those points where one
of the components of the second spatial derivative of some
user-selected quantities,
normalized to the square of the spatial gradient, exceeds some
pre-established 
threshold value. The quantities we check for refinement are
density, pressure and temperature. This default criterion is sufficient to
resolve the highly compressed regions on the scale of the clouds we
are interested in, so we do not add any additional, customized refinement criterion.

\section[]{Evolution of the cocoon}\label{ev_cocoon}
The injection of energy into the ISM/IGM can engender turbulence,
mostly because the jet is supersonic at and near the injection
point. Most previous work aimed at describing the general structure
and evolution of the cocoon, however, has paid more attention to the
global dynamics of the cocoon. Turbulence can have a significant impact on
the evolution of the embedded clouds, and for this reason we study it
in more detail in the next sections.

\subsection[]{Active phase: evolution during jet injection}\label{sec:evol:1}
\begin{figure*}
\centering
\includegraphics[scale=0.8,angle=90]{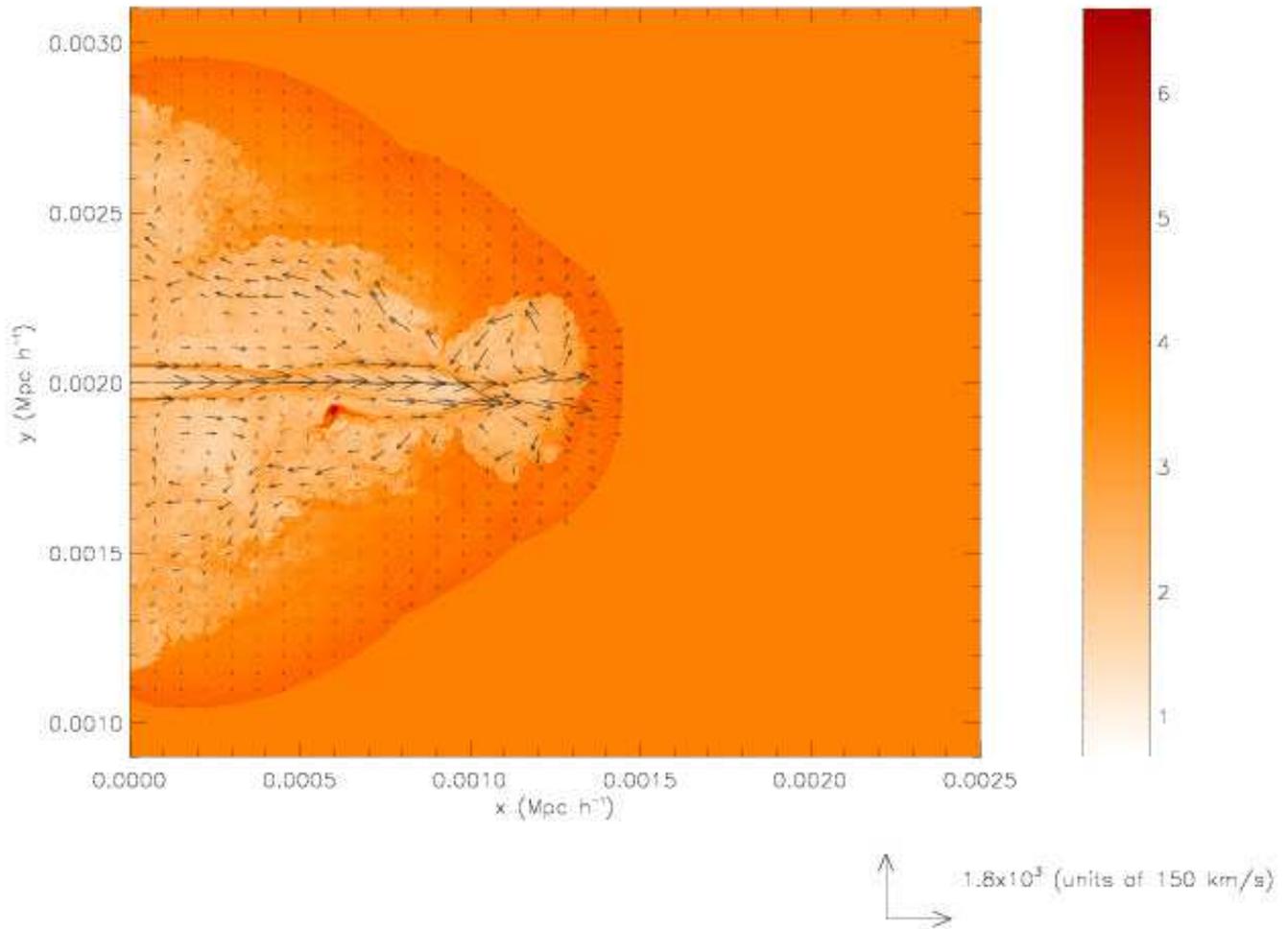}
\caption{Density and velocity distribution at $t=8.1\times 10^{4}$
  yrs. for run
  H4. The logarithmic density scale ranges from $10^{-3}$
  to 10$^{3}$
  e$^{-}$  cm$^{-3}$. The transition region between the cocoon and the
  external medium is threaded by a
  series of transonic shock waves (\emph{shocked ambient gas layer}). The high-density enhancement 
  within the cocoon originates out of the stripped material from a
  cloud initially set near (0.6,1.92) $h^{-1}$ kpc, which has been shocked by the jet.}
\label{fig_jet_h4_bw}
\end{figure*}

Soon after the jet enters the ISM/IGM a cavity forms, and the gas which
has been swept out piles up into a transition layer. In Figure~\ref{fig_jet_h4_bw} we show an output of
one of the runs at a rather advanced stage. We can easily distinguish an internal low-density, high temperature
{\textit cocoon} and a \textit{shocked ambient gas} region, externally
bounded by a tangential discontinuity with the outer ISM/IGM.
\noindent
One of the most interesting features we find is that these regions are also
\textit{dynamically} very different. The region containing the shocked gas is
threaded by a series of weak transonic shocks, and it has on average
an expansion motion, while in the cocoon a large-scale circulation
parallel and opposite to the main stream of the jet develops,
originating from gas reflected away from the region near 
the hot spot. This
circulation produces shear motions which then decay into weak turbulence
within the cocoon. Pre-existing clouds embedded within the ISM/IGM are
heavily affected by this turbulence.
\begin{figure*}
\centering
\includegraphics[scale=0.8,angle=90]{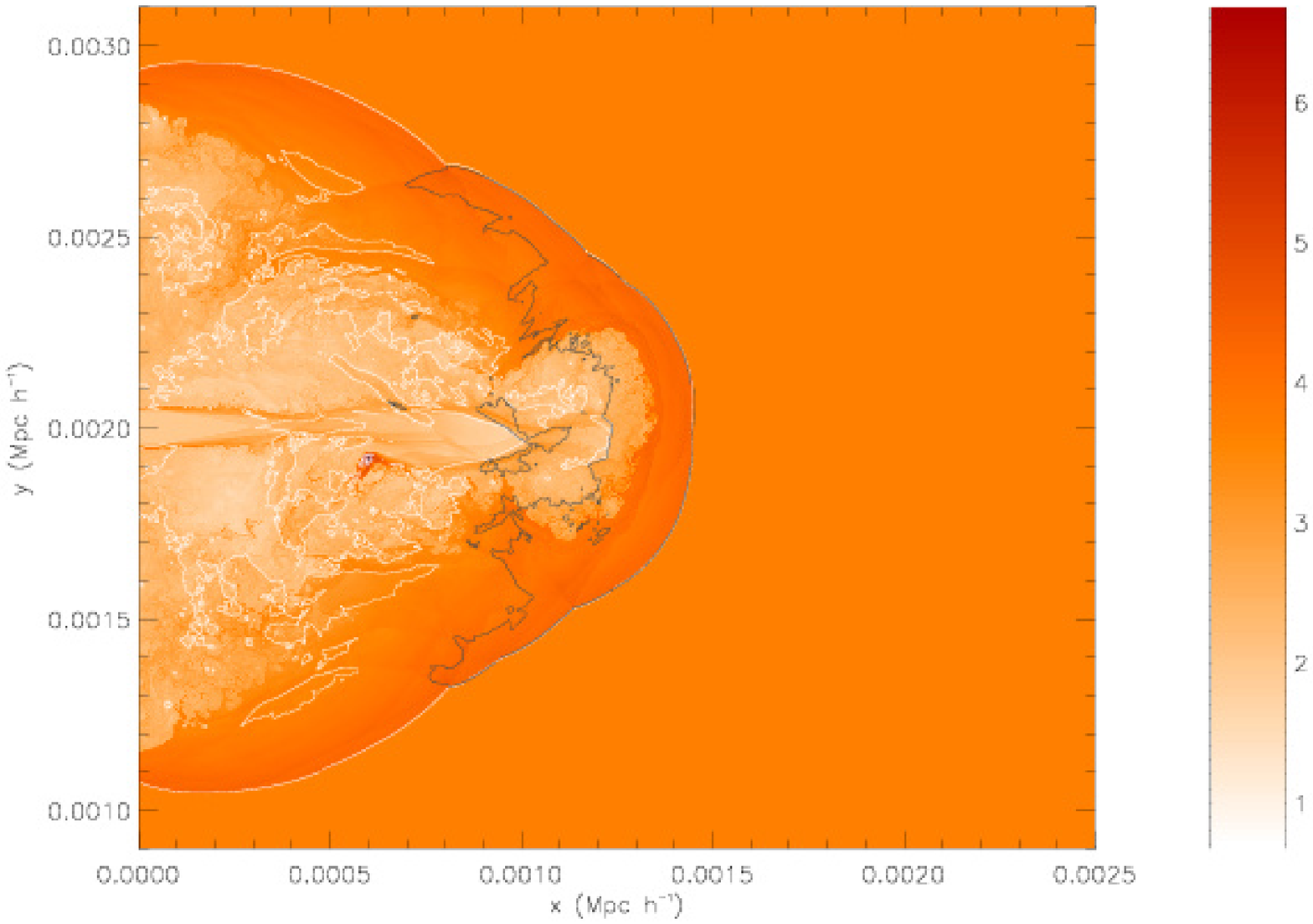}
\caption{Density and pressure distribution for the same output of
  Figure~\ref{fig_jet_h4_bw}. We show two pressure contours,
  corresponding to 5.77$\times 10^{9}$ (white) and 2.88$\times
  10^{10}$ (dark grey) K$\cdot {\rm cm}^{-3}$. }
\label{fig_jet_h4_bw_pres}
\end{figure*}
\noindent
The typical velocities of these shearing motions are large  but, due
to the very high temperatures within the cocoon (T$\approx
10^{9}-10^{11}\,$ K) , they are only moderately transonic. \\
As one can notice by inspecting Figure~\ref{fig_jet_h4_bw_pres}, the
pressure within the cocoon can reach high values, because the
temperatures are on average very high. Also the region around the
terminal part of the jet,
near the hot-spot, is subjected to high pressures, which increase
steadily until the cocoon's expansion is halted by the ram
pressure. We notice that the highest values of pressure are attained
within the 
{\em shocked ambient gas region}, mostly driven by the higher density, up to 3
orders of magnitude larger than the average density in the
cocoon. The
Mach number is higher near the jet, particularly near the injection
point, but in the overall region (cocoon and {\em shocked gas layer}) it never reaches values larger than $\mathcal{M} \approx
3.-3.5$. The dynamical evolution of the hot spot, i.e. the region
between the tip of the jet and the terminal part of the cocoon, is
quite interesting. Here, however, we will concentrate on those features
more directly related to the interaction with clouds,
leaving to further work a more detailed analysis of the features of
interest relevant to the modeling of the radio jet.
\subsection[]{Passive evolution}
All runs were stopped when the jet was still active, except for run H4,
which was continued for about $5\times 10^{6}\,$ yrs. after the jet
was switched off. This time was chosen to be $t \approx 10^{7}\,$
yrs., within the range of a typical duty cycle for the AGN.
\begin{figure*}
\centering
\includegraphics[scale=0.7,angle=0]{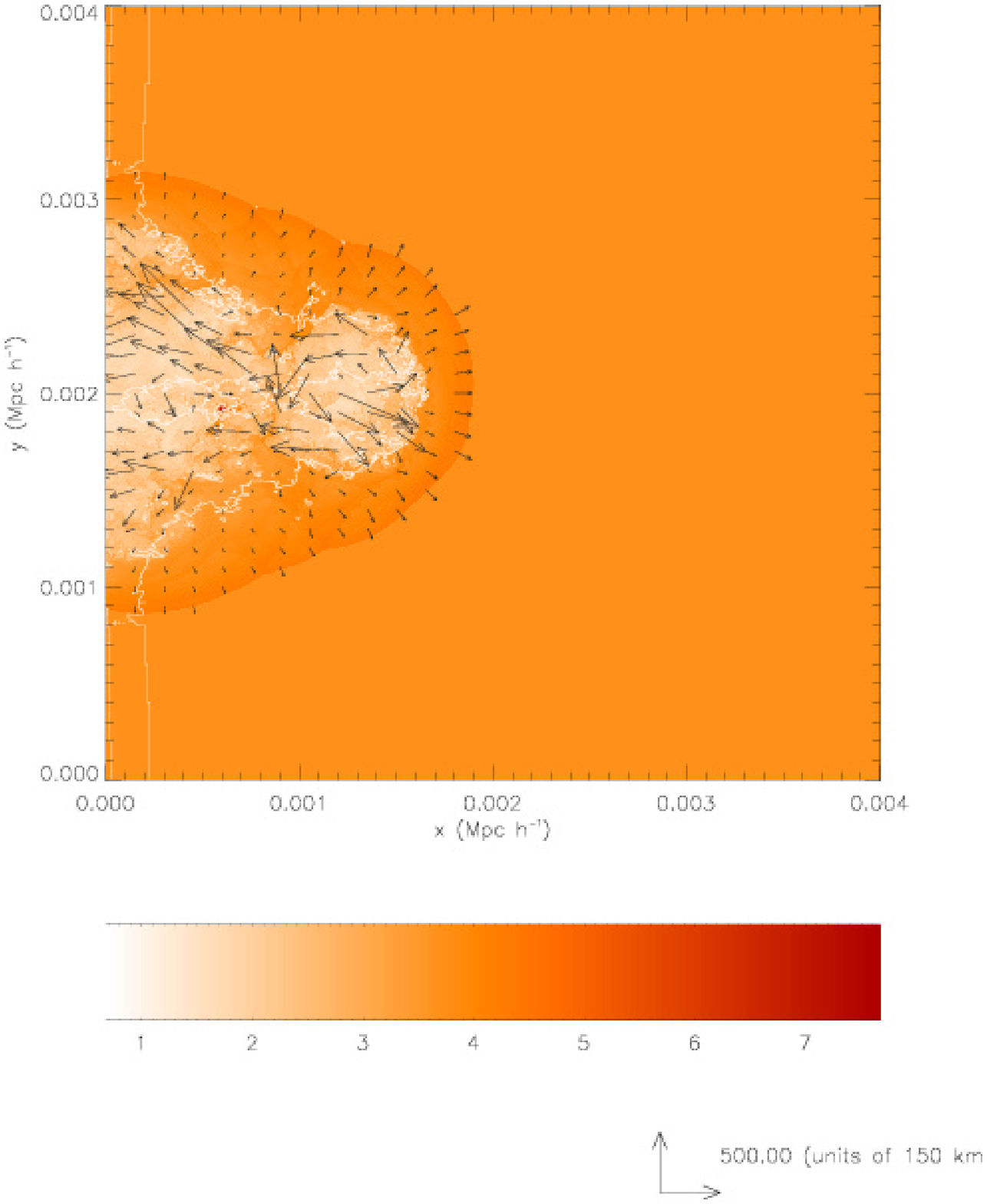}
\caption{Density field of run H4 at $\Delta t=2.45\times 10^{5}$ yrs after the
  jet has been shut off (i.e. at a time $t=1.00245\times 10^{7}$ yrs
  since the beginning of the simulation). The cocoon and the tangential discontinuity
  are now dynamically very different, and their boundary is marked by
  a zero velocity contour.}
\label{fig_ev_sfr1}
\end{figure*}
The cocoon expands up to a maximum radius determined by the jet power
and the environmental ram pressure. When the duty cycle of the AGN is
completed, the jet injected power decreases very rapidly, and the
cocoon is not anymore fed by the jet. The
evolution is mostly driven by inertial motions: a snapshot is
shown in Figure~\ref{fig_ev_sfr1}. The typical velocities are still
quite high within the cocoon, but the expansion of the shocked ambient gas 
region has already been slowed down by the ISM/IGM ram pressure, and
it is now decelerating, while slowly expanding. Notice that the cloud
is threaded by arc-like internal shocks within the cloud
(Figure~\ref{fig_ev_sfr4}, {\em left}), previously
noticed in similar simulations of shock-cloud interactions \citep{2006ApJS..164..477N}. \\ 
\begin{figure*}
\centering
\includegraphics[scale=0.7,angle=90]{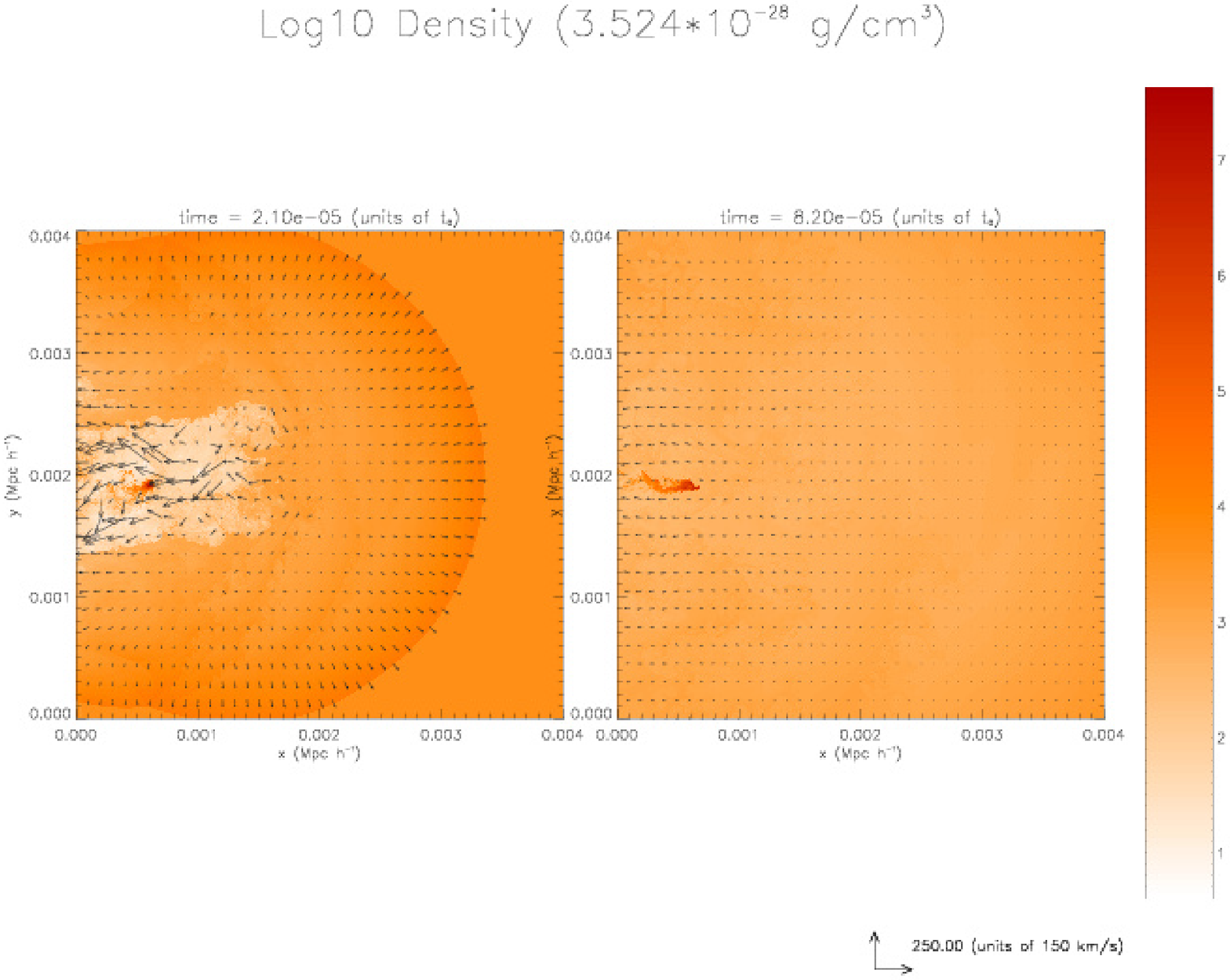}
\caption{Density evolution at two different epochs after the jet
has been switched off. The system evolves towards homogeneity and
equilibrium on a temporal scale mostly determined by the decay of the
inertial motions. Time is measured in units of Hubble time $t_{0}$, so
the plot on the left is at an epoch $t=2.835\times 10^{5}\,$ yrs., and
the one on the right plot for $t = 1.107\times 10^{6}\,$ yrs.}
\label{fig_ev_sfr2}
\end{figure*}
\noindent
\noindent
The further evolution of the cloud during the passive phase shows some features not emphasized in previous
papers. A continuous, decelerating flow now takes place
within the cocoon, primarily driven by the inertial motions, because
at the typical densities of the cocoon ($n_{e} \approx 10^{-4} -
10^{-2}\, {\rm cm}^{-3}\,$) the gravitational pull is less than the
inertial force. A zero total velocity line now separates the cocoon
from the external ISM. In the latter the decelerating, outward 
directed expansion motion changes to a wind as the plasma leaks out of
the simulation box and the density decreases. A similar motion, but
directed in the opposite direction, drives the gas past the left
vertical boundary, i.e. towards the AGN and the galaxy bulge,
increasing the density within this region. As a consequence of the
generalized density decrease, the cooling time increases even further,
although it was already much larger than the free fall time-scale.\\
%
\noindent
This back-flow within the cocoon has some interesting consequences for
the embedded clouds. During the active phase, the cloud was already
subject to a significant compression from the turbulent motions, and,
as a consequence, its density had also risen. Its temperature
 was
however typically higher than $5\times 10^{5}-10^{6}\, {\rm K}$, a
level maintained mostly by the turbulent dissipation. However, during the
passive phase, the cloud is embedded within a continuous decelerating
stream, which is now more laminar than turbulent
(Figure~\ref{fig_ev_sfr2}). This flow continues to compress the cloud,
which cools efficiently: its average density also increases, and its temperature decreases, a
trend which can be clearly seen on the right hand side of
figure~\ref{fig_ev_sfr4}. On a longer time scale the
cloud eventually is completely stripped, due to KH instabilities.
\begin{figure*}
\centering
\includegraphics[scale=0.7,angle=90]{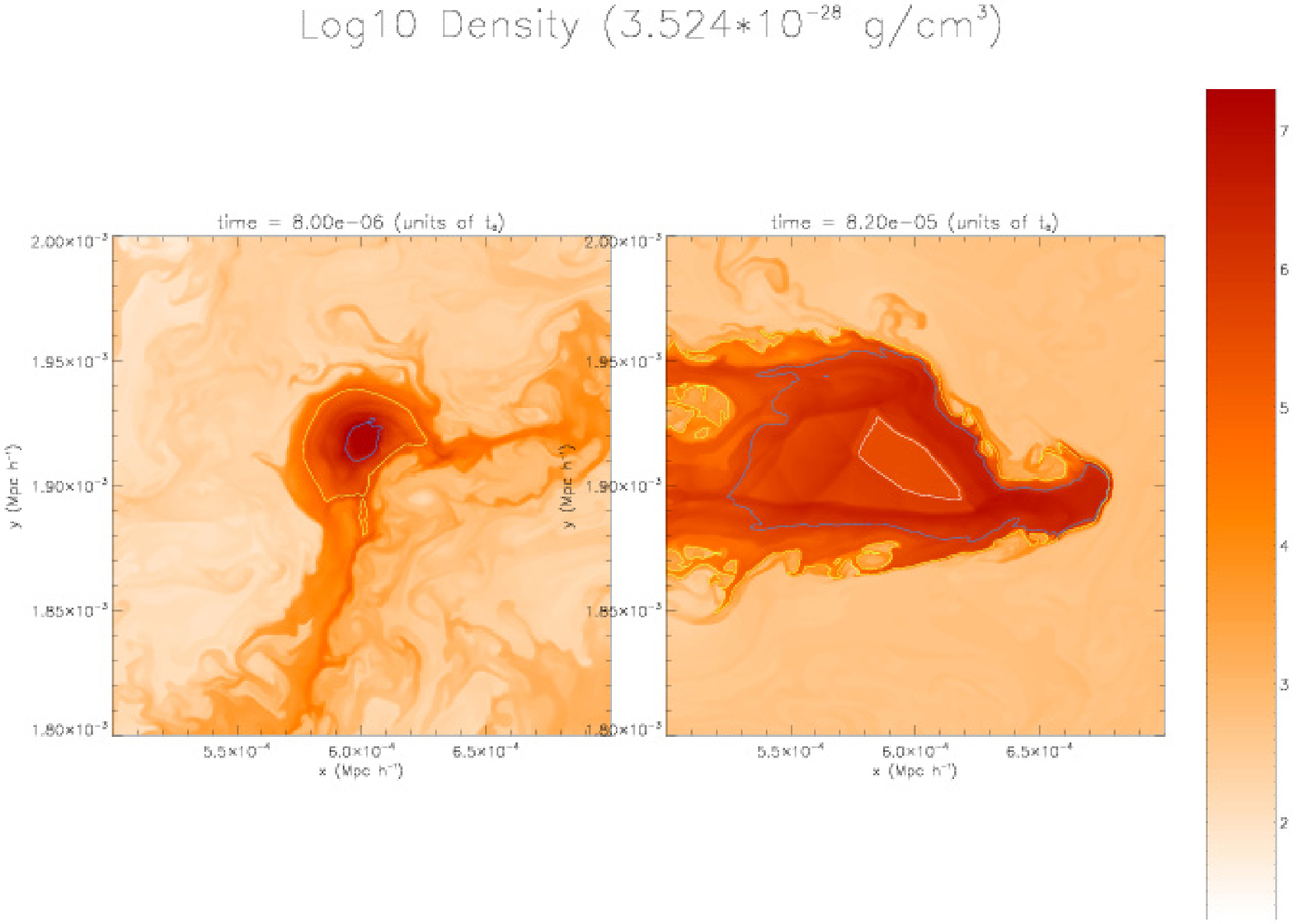}
\caption{Evolution of density and temperature in the compressed
  cloud, during the active phase ({\em left}), during the which the
  jet is feeding energy into the cocoon, and some time after it has
  been switched off ({\em right}). The contours superimposed on the density maps trace regions of
  equal temperature, and correspond (from the inner- to the outermost
  contour) to the following temperatures: ${\rm T} = 10^{4}, 10^{7}
  {\rm and} 10^{9}\, $K.}
\label{fig_ev_sfr4}
\end{figure*}
Our simulations do not have
the spatial and temporal resolution to further follow the
fragmentation of this cloud, which will be treated in a separate paper. The overall action of the cloud's
compression during the \textit{active} jet phase, and of the cooling
and further compression during the passive phase, can result into an
occasional enhancement of the region of the cloud where favorable
conditions for star formation are possible, as can be seen from
figure~\ref{fig_ev_sfr4}. The gas stripped away by KH instabilities
from the cloud tends to form filamentary, high density structures,
where temperatures can reach high values (${\rm T} \sim 10^{4} -
4\times 10^{5}\,
$K). Star formation within these filaments, due to these high
temperatures, is thus inhibited, and the cloud eventually is
completely destroyed. 
\section{Probability density distribution}\label{ev_pdf}
The cocoon represents for the cloud an environment radically different
from the hot ISM: on average, temperatures are higher and densities
lower, and it is also in a turbulent state, permeated by a series of
turbulent eddies, which produce a random, intermittent series of
stresses on the cloud. This scenario is more complex than those
envisaged in previous models of cloud evolution under external shocks
\citep{1994ApJ...420..213K,2006ApJS..164..477N}, and a model of the turbulence within the
cocoon is then a prerequisite to developing a model of the evolution of
clouds within a turbulent environment. We will develop the latter in a
following paper: here we will only consider the density PDF within the
cocoon.\\ 
Simulations of turbulence in the ISM suggest that, on galactic scales, 
a lognormal Probability Distribution Function (LNPDF) provides a
viable description of the distribution of density fluctuations over a large range
of densities typical of galactic ISM 
\citep[e.g.][]{1997ApJ...474..730P, 2007ApJ...660..276W, 2001ApJ...547..172W}. The physical 
conditions
of the system considered in this work, however, are very different from those
typical of the galactic ISM, where temperatures are
considerably lower (T$\approx 10-10^{2}\,$ K) and densities higher, so
we do not expect \textit{ a priori} to 
find the same PDF. Moreover the presence of the jet, which injects
energy and momentum into an expanding cocoon, results in a
\textit{non-stationary} background. 
\begin{figure}
\centering
\includegraphics[scale=.4,angle=0]{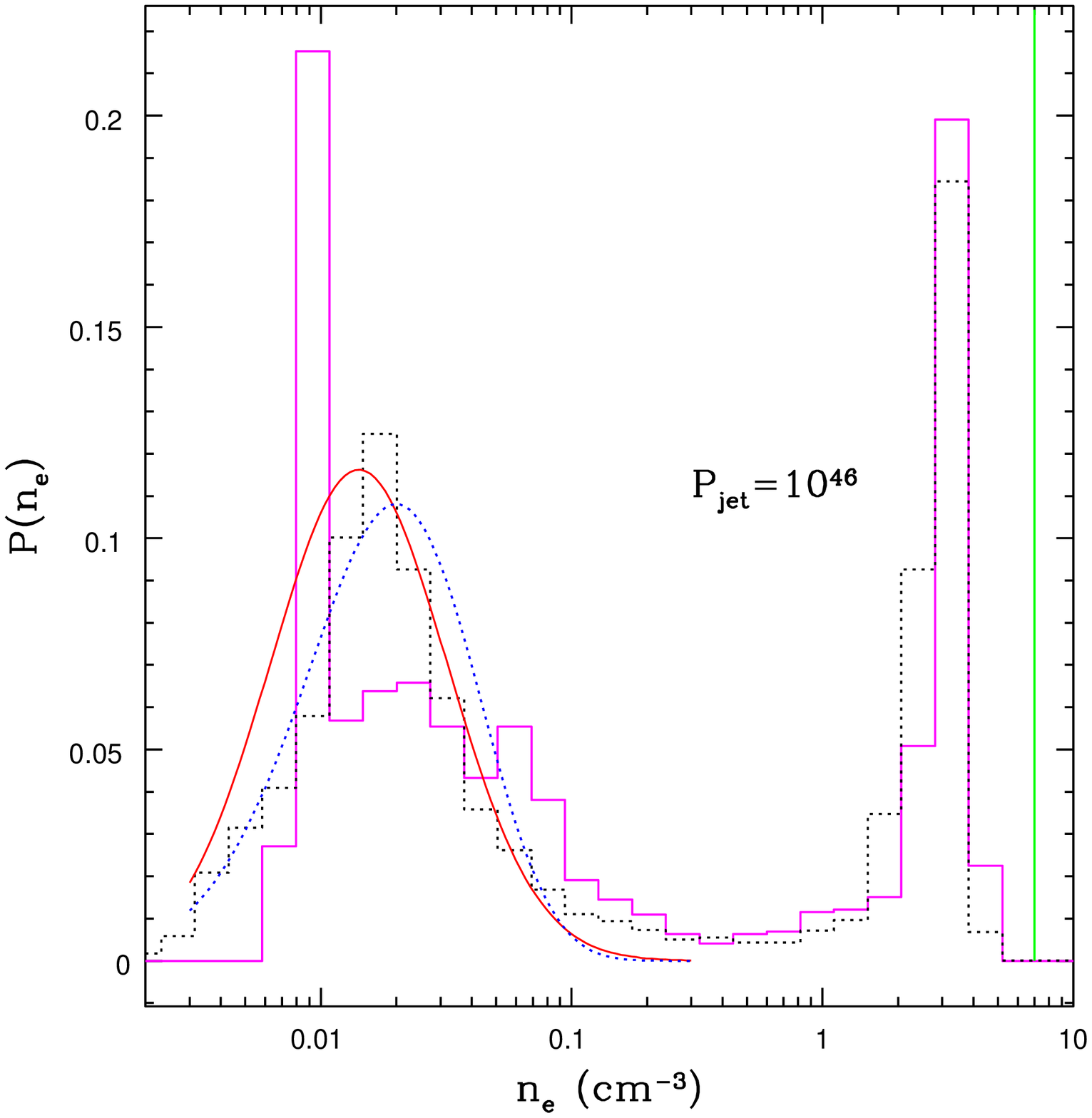}
\caption{Density PDF for run H4. The analysis was performed only
  within the cocoon/shocked ambient gas region. The two histograms show the PDF at
  two different epochs: t=$1.35\times10^{4}\,$ yrs. (dotted), and t=$9.45\times
  10^{4}\,$ yrs. (continuous). The curves show two best fits for the
  latter epoch: am exponentially truncated power law {\em continuous}
  and a a modified lognormal ({\em dotted}). The vertical line at
  $n_{e} = 7$ is the maximum density possible for fully radiative
  shocks \citep{2000ApJS..127..245B}. Best fit parameters are given
  by: (A,b,n) = ($1.141\times 10^5, 14.486, 24.612$)
  (eq.~\ref{eq:pdf:1}), and: (B,$\langle x\rangle,\sigma$, b) =
  ($3.66\times 10^{-3}, -1.6,0.498, 2.005$) (see eqs.~\ref{eq:pdf:1}
  and ~\ref{eq:pdf:2}).}
\label{fig_pdf_h4}
\end{figure}
In Figure~\ref{fig_pdf_h4} we show the evolution of the density PDF
for one of the runs (H4). One notices that the PDF is bimodal, with
two peaks at densities respectively lower and higher than the 
initial ambient density (\nel=1 ${\rm cm}^{-3}$). These two peaks correspond to two
  spatially separated regions: the low density distribution arises
  predominantly within the cocoon, while the high density component
  is associated with the gas in the shocked gas region. The continuous
  and dotted curves
  are two least-square fits of the low-density PDF, with two
  different fitting functions: an exponentially truncated power law (continuous curve),
\be
P_{tr} = A\mid x\mid^{n}\exp\left(-\frac{x}{b}\right)
\label{eq:pdf:1}
\ee
 and a modified lognormal (dotted curve):
\be
P_{lm} = B\exp\left[-\frac{\left(x - \langle
 x\rangle)\right)^{2}}{\sigma^{2}} - bx\right]
\label{eq:pdf:2}
\ee
where: $x={\rm Log}(\rho)$. 
The quality of the
two fits is comparable: $\chi^{2}/d.o.f = 1.76$ (truncated power law)
and $\chi^{2}/d.o.f = 1.32$ (modified lognormal). A modified lognormal
 distribution is generally expected  in any fluid system whose density distribution
  is the result of a series of uncorrelated shocks
  \citep[e.g][]{1994ApJ...423..681V,1997ApJ...474..730P,1998PhRvE..58.4501P,2007arXiv0706.0739K}.
  Such a PDF has also been found in simulations of
  turbulence and global star formation in galactic discs
  \citep{2001ApJ...547..172W,2006ApJ...641..878T,2007ApJ...660..276W},
  thus suggesting that it could also arise in stationary systems
  where turbulence is not only artificially forced within the
  simulation box.  Note that in the model of
  \citet{1998PhRvE..58.4501P} this particular form for the PDF should
  be independent of the dimensionality of the system, although the
 parameters characterizing the distribution will depend on it.
It seems reasonable to assume that the main driver of this weak 
  turbulence are multiple shocks within the cocoon, forming initially
  from the evolution of Kelvin-Helmholtz instabilities at the jet-ISM 
  interface. These shocks then propagate within the cocoon, and are
  reflected at the internal interface with the shocked gas
 region. 

\section[]{Evolution of star formation rate}\label{jet_cloud}
\begin{figure}
\centering
\includegraphics[scale=0.4,angle=0]{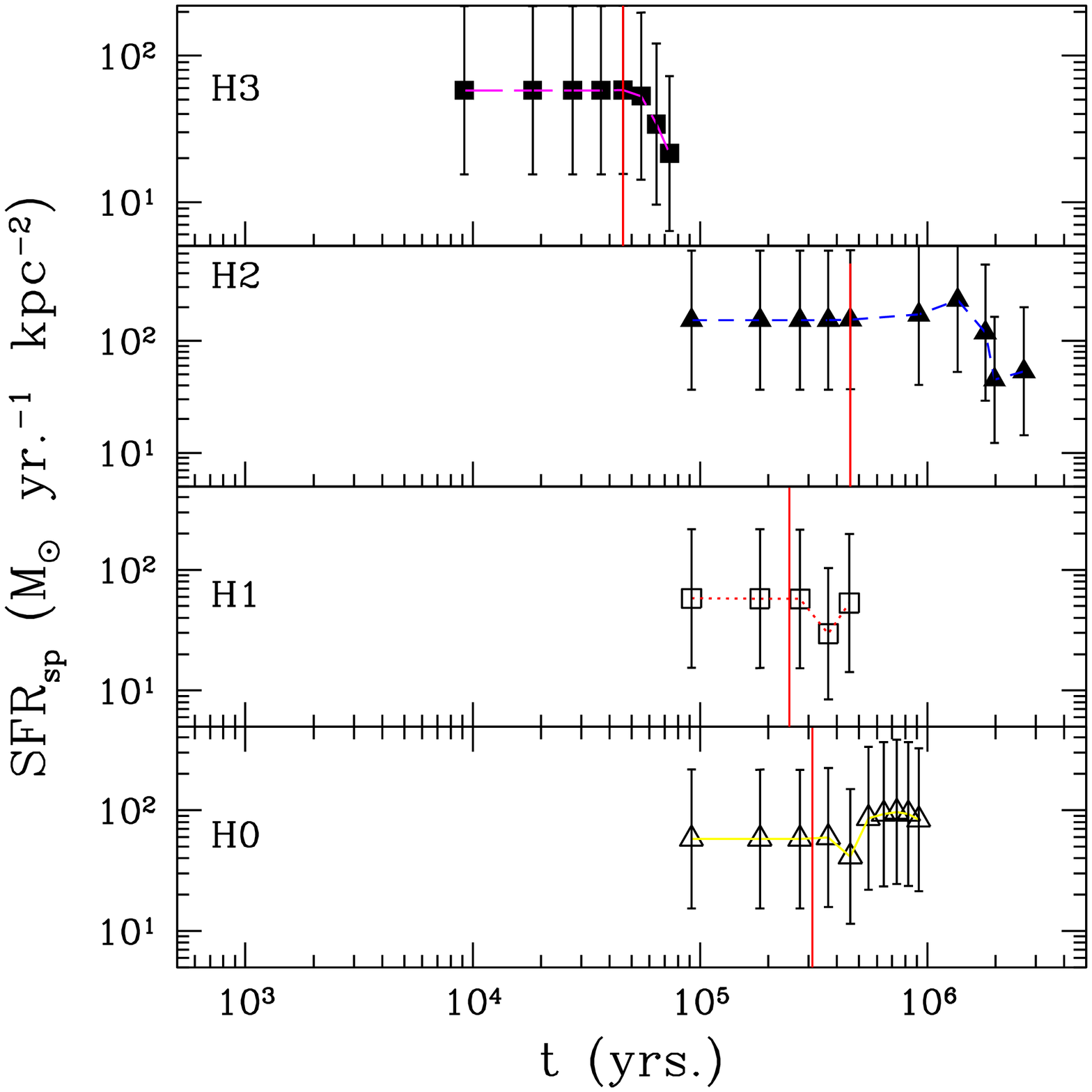}
\includegraphics[scale=0.4,angle=0]{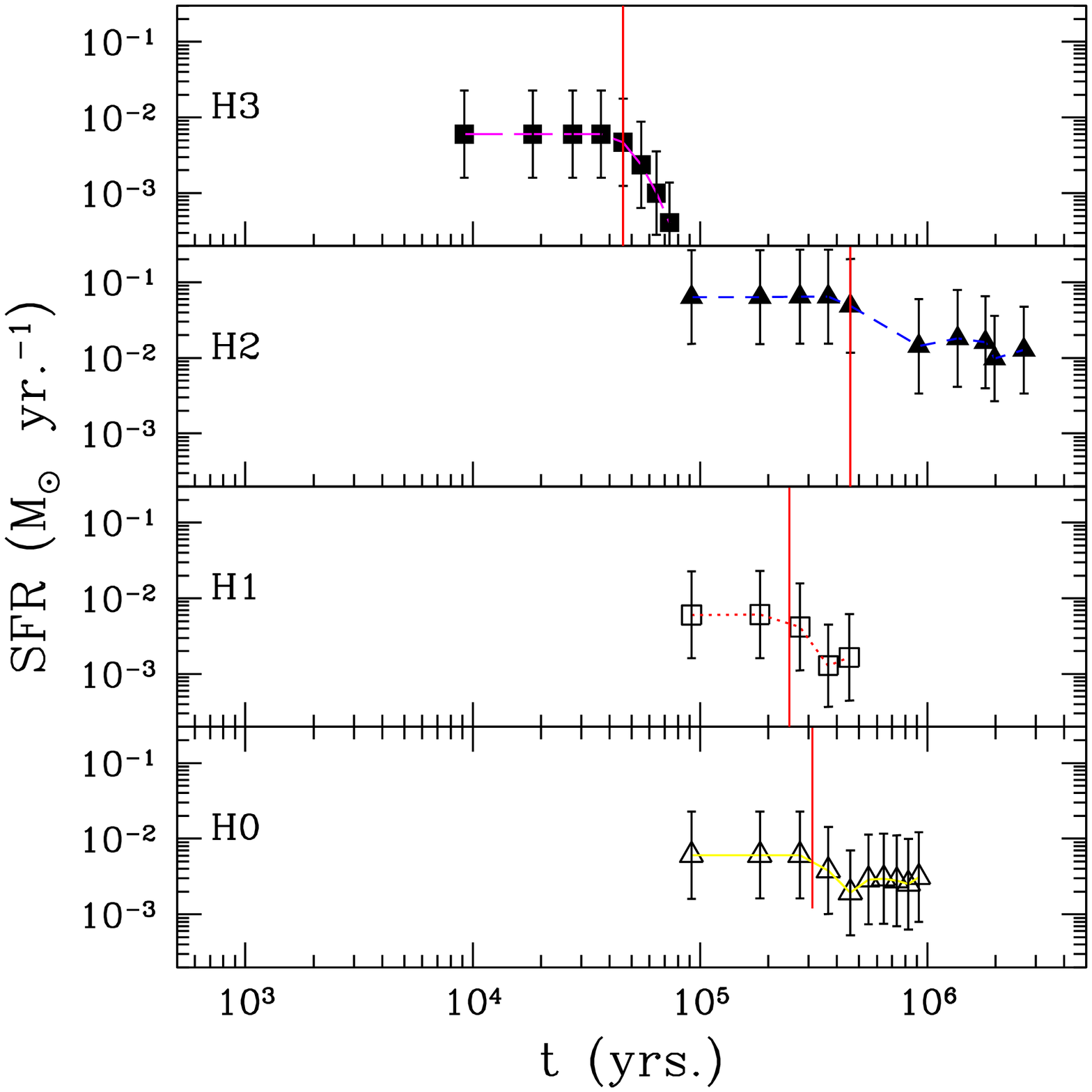}
\caption{Evolution of the cloud's star formation. The red vertical bar
labels the time when the jet/cocoon first hits the cloud. Error bars
are computed from the $\pm 1 \sigma$ errors of the Schmidt-Kennicutt
law (Kennicutt 1998). \emph{Upper plot}: Specific star formation rate,
\emph{lower plot}: Total star formation rate over the whole cloud.}
\label{fig4}
\end{figure}

\begin{figure}
\centering
\includegraphics[scale=0.4,angle=0]{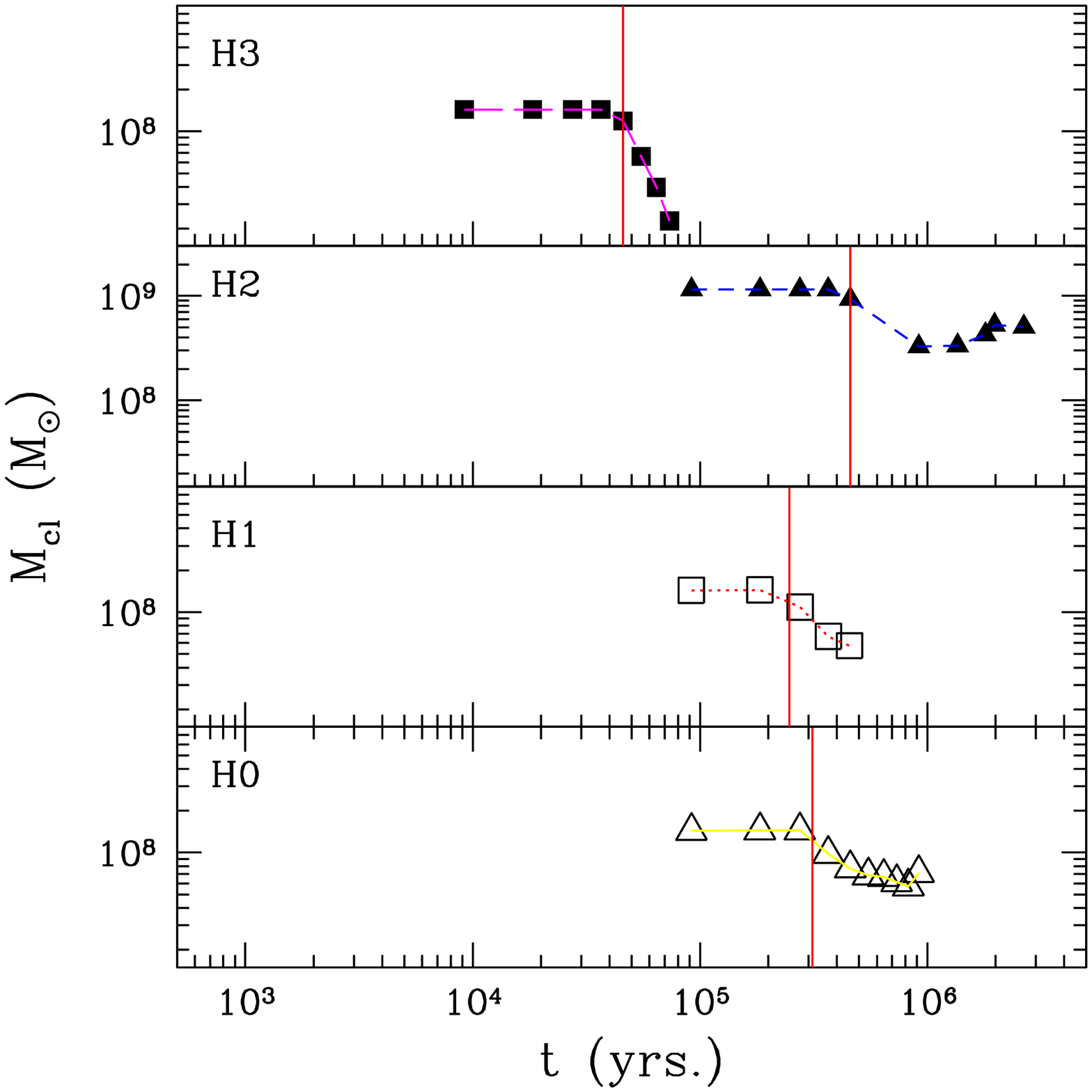}
\caption{Evolution of the mass of the cloud's star-forming region.}
\label{fig_mass}
\end{figure}

\begin{figure}
\centering
\includegraphics[scale=0.4,angle=0]{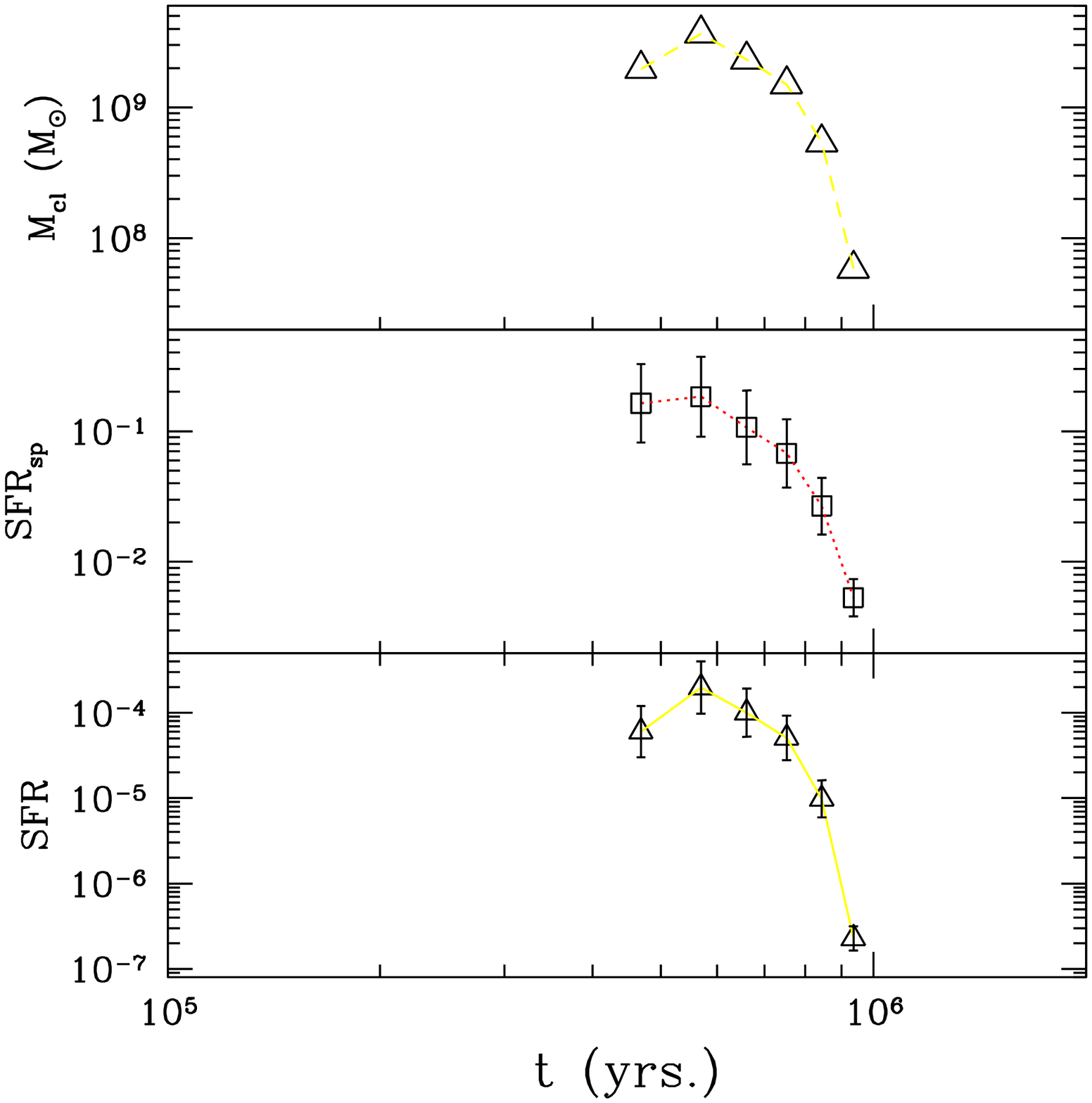}
\caption{Mass and SFR evolution after the switch off of the jet (run
  H4). Time is now measured in yrs. after the switch off epoch
  (arbitrarily fixed at t=$10^{7}$). }
\label{fig_swoff}
\end{figure}

The influence of the jet/cocoon on star formation within the cloud is
the result of the competition between different physical factors. The
impact of the shock and the increased pressure within the cocoon
result in a overall compression of the cloud, which tends to {\em increase} its
density and specific star formation rate. On the other hand, the
increase of temperature and stripping of the outer regions due to KH
and other instabilities tend to reduce the mass of the cloud, thus
contributing to a {\em decrease} of the global star formation rate.\\
The detailed temporal evolution of the cloud is governed by the
turbulence within the cocoon, and subsequently by the evolution of the 
back-flow. In Figure~\ref{fig4} we show the evolution of the {\em
  specific} and {\em global} star formation rates during the active
phase, when the jet is still feeding energy into the cocoon. One notices that in runs H0-H2 the specific SFR
occasionally increases, due to the increase of the average density of
the star forming region. The {\em global} SFR  also shows similar
fluctuations, although of smaller amplitude, in all these runs. 
A similar trend is observed for the evolution of the mass of the star
forming region, Figure~\ref{fig_mass}: some episodic increases are
seen, in temporal and sign correspondence with the fluctuations of the
SFRs. However, stripping tends to diminish the mass of the cloud, so we are
forced to conclude that these episodic increase of the star-forming
region can only be attributed to occasional cooling, driven by thermal
instability. This gas is however soon warmed up again by the warm, turbulent
environment of the cocoon, and the SFR then decreases.\\
A similar trend is observed in run H4, during the passive phase. We have already seen from
Figure~\ref{fig_ev_sfr4} that the cloud, exposed to the back-flow from
the jet, becomes more filamentary and cold. However, we see from
Figure~\ref{fig_swoff} that its mass decreases dramatically, because
the inner cold regions become less dense, and consequently both the
specific and the global SFRs diminish. 
We can thus conclude that, in general, the interaction with the jet/cocoon
tends to inhibit in the long term star formation within the
cloud. This suppression however is not continuous, but seems rather to
proceed in an intermittent way, and is interrupted by short episodes
during which the SFR occasionally increases. This trend seems to
continue during the passive phase, when the cloud is completely
stripped and its SFR decreases rapidly.

\section{Discussion}\label{sec_disc}
Some features of the evolution of the clouds described in this work
have been previously found in earlier work. Our initial setup is
very similar to that used by \citet{2002A&A...395L..13M}: the initial
temperatures of the ISM and clouds are the same, and our clouds are
also in pressure equilibrium with the diffuse phase. We also find that
the stripped cloud seems to fragment into filamentary structures, whose
density and specific SFR seem to occasionally increase. Although these
structures seem to persist, we have shown that their global star
formation rate tends to decrease.\\
Recent observational evidence suggests the star formation history in 
early-type galaxies can be more complex than in the
standard, monolithic scheme. There is both evidence of quenching of star
formation \citep{2007arXiv0707.3570K,2006Natur.442..888S} and of
recent star formation episodes
\citep{2007arXiv0709.1394S,2007arXiv0709.0806K,2007IAUS..241..546T,2007RMxAC..28..109Y}:
both  could be induced by AGN activity. It is then important to
understand the physical mechanism of jet-ISM interaction, and how star
formation can be affected. The simulations we have presented in this
work represent an attempt to introduce some realistic physical
ingredients of the ISM structure into a numerical model, and to
understand how these affect the evolution of SFR in the cloud\\
\noindent
There are some features which have not yet been included into this
model. One  is related to the quantitative relevance of this
model. Although the duration and frequency of AGN jets are not well
known \citep{2002NewAR..46...75B}, typical estimates of the
\textit{active} phase of the jet range between $10^{7} - 10^{8}$ yrs,
comparable to those adopted in the simulations presented in the
previous sections. Empirical evidence from 
  modeling of SMBH feeding in the context of hierarchical structure
  formation
suggests that the AGN phenomenon persists for several Gyr
  \citep{2005MNRAS.359.1363M,2004MNRAS.350..456B,2004cbhg.symp..405H},
  or many galaxy dynamical times. Hence AGNs are certainly capable of
  providing significant feedback over the time-scales relevant to
  star formation and gas accretion in hierarchical galaxy  formation models.

Our simulations are restricted to only \textit{one
  episode} of AGN activity: a further episode of jet injection would
certainly propagate into a very hot and low-density ISM, because our
simulations show that the diffuse ISM/IGM within the cocoon reaches
very high temperatures (T$\approx
10^{8}-2\times 10^{11}\,$K) and low densities ($n_{e} \approx
10^{-2}-10^{-1}\, {\rm cm}^{-3}$), depending on the initial temperature and on
the jet power at the time of its maximum expansion. Under these
conditions, the cooling time exceeds the dynamical time by a factor:
$t_{cool}/t_{dyn}\approx 6.5\times 10^{2} - 3\times 10^{5}$, thus the
heated gas does not cool significantly (as already noted by
\citealt{2001ApJ...562..618I}). Thus, a second jet would propagate into
an environment where the ram pressure will be 
considerably higher, $p_{ism} \approx 10^{6} - 2\times 10^{10}\,
  {\rm K}{cm}^{-3}\,$ (the initial pressure in the ISM was $p_{ism} = 10^{7}\, k^{-1}$), and would probably be quenched. We
will verify these predictions in subsequent work.\\
Despite the very high temperature within the cocoon, as soon as the
jet is switched off, we observe that the shocked clouds which
are not evaporated are subject to a back-flow originating from the gas
within the cocoon. As a consequence, they become filamentary, and
their temperature decreases, but we have seen that the mass of the
  star-forming regions within the cloud also diminishes, so
  eventually the global SFR is reduced. In a subsequent paper,
we will present results of a more realistic simulation where we
consider the propagation of a jet into a \textit{multicloud} system,
and we will then be able to address the problem of AGN
feedback in a more quantitative, statistical fashion.\\ 
Our  simulations differ significantly
from those of KA07, because the turbulence
within the cocoon surrounding the propagating jet is generated
naturally by the interaction of the jet with the ISM/IGM, while KA07
studied the turbulence associated with the disruption of pre-existing
ISM/IGM clouds due to Kelvin-Helmholtz instabilities at the jet-IGM
interface. There are however some features which we have not addressed
in our simulations. \citet{2007arXiv0707.3668S} have
simulated the propagation of a jet which first emerges out of a
gaseous disc. They note that the disc has a strong effect on the
energy budget of the jet and also on its morphology: during its
initial propagation the jet percolates through the disc, and it needs
more time to overcome this initial pressure and emerge out of the disc
into the ISM. Sutherland \& Bicknell do not consider the compression
of pre-existing clouds: the main target of their work is a detailed
study of the morphology of the emerging jet, and a comparison with
observations. A significant gaseous component, often distributed in
disc-like structures, seems however to be present  in relevant
amounts ($10^{9}-10^{10} \textrm{M}_{\sun}$)  also in a
fraction of early-type galaxies
\citep{2007MNRAS.377.1795C,2006MNRAS.371..157M,
  2007arXiv0709.1394S,1996A&AS..120..463M,1994A&AS..105..341G}. 

\section{Conclusion}\label{sec_conc}
The simulations we have performed in this work have as a target the
propagation of the jet into the inhomogeneous ISM of an host galaxy. We
have restricted our interest to the central region of a model
elliptical galaxy ($r < 4\,$ $h^{-1}$ kpc), and we have studied the
propagation of the jet and its interaction with a single cloud. The
adoption of an AMR code like FLASH has proved crucial for studying,
in the same simulation, both the large-scale properties of the jet-ISM
interaction and the interaction with a small cloud. We have modeled
star formation within a small, dense cloud assuming that it follows
the Schmidt-Kennicutt law, and we have studied how the SFR is modified
by the impact of the jet/cocoon. In general, we find that the SFR could be
occasionally enhanced, but in the long run the cloud is stripped by KH
instabilities and its SFR decreases. After the jet has been switched
off, a laminar back-flow flow develops which continues to compress and
strip the cloud, until the latter loses a significant fraction of its
mass. Due to the very long cooling time of the cocoon, the cloud is
embedded in a very hot, low-density medium, and this eventually results
in suppression of star formation. We note that our results are
preliminary being restricted to simulations in 2D for a single
cloud. In further work we will extend our study to full 3D simulations
and to an inhomogeneous protogalaxy. 
\section*{Acknowledgments}
The work of V.A.-D. has been supported by the European Commission,
under the VI Framework Program for Research \& Development, Action
``{\em Transfer of Knowledge}'' contract
MTKD-CT-002995 (''{\em Cosmology and Computational Astrophysics at
  Catania Astrophysical Observatory}''). V.A.-D. would also express
his gratitude to the staff of the subdepartment of Astrophysics,
Department of Physics, University of Oxford, and of the Beecroft
Institute for Particle Astrophysics, for the kind hospitality during
the completion of this work.\\
The software used in this
work was partly developed by the
DOE-supported ASC/Alliance Center for Astrophysical Thermonuclear
Flashes at the University of Chicago. Finally, 
this work makes use of results produced by the PI2S2 Project managed
by the Consorzio COMETA, a project co-funded by the Italian Ministry
of University and Research (MIUR) within the {\em Piano Operativo Nazionale 
"Ricerca Scientifica, Sviluppo Tecnologico, Alta Formazione" (PON
2000-2006)}. More information is available at http://www.pi2s2.it (in
italian) and http://www.trigrid.it/pbeng/engindex.php.


%
\bibliographystyle{mn2e}
\bibliography{biblio}

\appendix

\section[]{Pressure confined clouds} \label{append_1}
The truncated spherical equilibrium configurations
studied by \citet{1999MNRAS.307..203S} provide a convenient model for
IGM clouds, because they contain two key physical ingredients: they are
isothermal and confined by the pressure of an external medium. These
configurations are described by solutions of a dimensionless Lane-Emden equation:
\be
\frac{d}{d\zeta}\left[\zeta^{2}\frac{d\left(\ln\tilde{\rho}\right)}{d\zeta}\right]
= -\tilde{\rho}\zeta^{2} 
\label{eq_app_a1}
\ee
Here $\zeta = r/r_{0}$ and \rhot$= \rho/\rho_{0}$ are defined as
dimensionless distance and density, respectively. The characteristic
scales are related by:
\be
r_{0} = \frac{\sigma}{\left( 4\pi G\rho_{0}\right)^{1/2}} = \left(\frac{k_{B}T_{cl}}{4\pi G\rho m_{p}\rho_{0}}\right)^{1/2}
 \label{eq_app_a2}
\ee
where we have introduced the definition: $\sigma =
(k_{B}T_{cl}/m_{p})$, $T_{cl}$ being the cloud's temperature. The TIS models studied by
\citet{1999MNRAS.307..203S} are exact solutions of the initial value
problem for eq.~\ref{eq_app_a1} subject to two initial conditions at the origin: 
\be
\rho(\zeta=0) = 1, \hspace*{0.5cm} \frac{d\rho}{d\zeta}\mid_{\zeta=0} = 0
 \label{eq_app_a3}
\ee
The first condition enforces the initial
value for the density, while the second one selects those solutions
which are not singular, i.e. it defines a \textit{core}. These 
configurations still have infinitely extending density profiles
decaying as $r^{-2}$ for $r \ll r_{0}$, but in presence of an
external pressure the profiles get truncated at a radius \rtr \citep{1957MNRAS.117..562M}. This external
confining pressure \ptr is then an additional parameter, which will
determine the dimensionless truncation distance
$\zeta_{t}$. Thus, these models are completely characterized by three
parameters: \ptr, the total cloud mass \mtot and radius \rtr. From
these we can deduce the characteristic density $\rho_{0}$ and scale
$r_{0}$, by assuming that clouds are described as TIS profiles.
\begin{figure}
\centering
\includegraphics[scale=0.4]{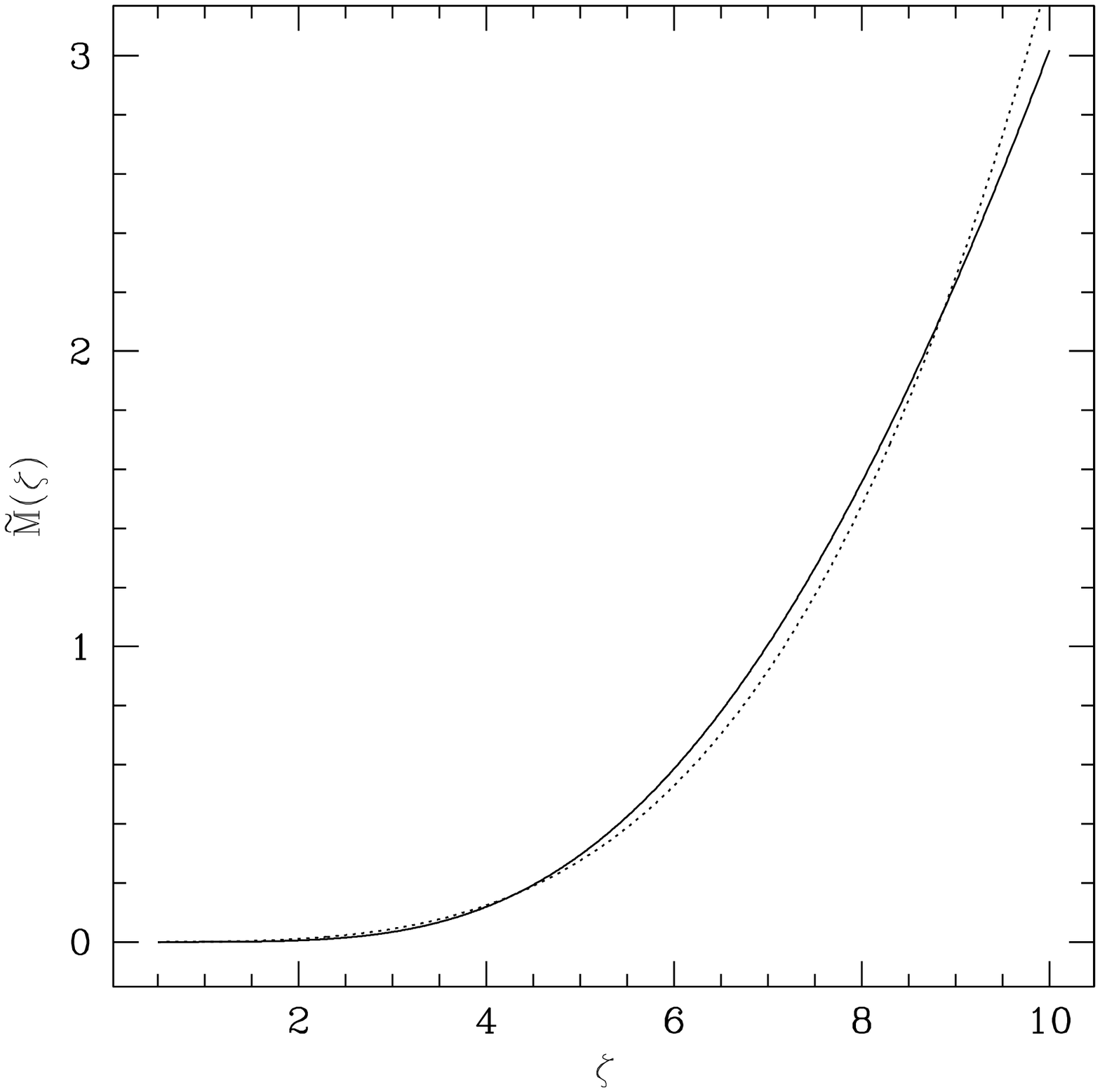}
\caption{Power-law approximation to the TIS mass profile. The
  continuous curve is the function $\tilde{M}(\zeta)$, the dotted one
  the approximation eq.~\ref{eq_app_a6}.}
\label{fig18}
\end{figure}
From eq. (35) of \citet{1999MNRAS.307..203S} \mtot is expressed in
terms of the averaged mass as:
\be
M_{t} =
\left(\frac{k_{B}T_{cl}}{Gm_{p}}\right)^{3/2}\frac{\tilde{M}(\zeta_{t})}{\left(4\pi\rho_{0}\right)^{1/2}}
 \label{eq_app_a4}
\ee 
where: $\zeta_{t} = r_{t}/r_{0}$ and we have defined:
\be
\tilde{M}(\zeta) = \int_{0}^{\zeta}ds s^{2}\tilde{\rho}(s)
 \label{eq_app_a5}
\ee
In the interval $0\leq\zeta\leq 9$ the following expression provides
a good approximation to eq.~\ref{eq_app_a5} with the density
$\tilde{\rho}(s)$ specified by the TIS solution:
\be
\tilde{M}(\zeta) = \lambda \zeta_{t}^{1/\beta}
 \label{eq_app_a6}
\ee
where the coefficient $\beta = 0.28\pm 0.04$ has been introduced in
eq.~\ref{eq:cleq:2}. This fit is accurate to 4\%, as can also be seen from
Fig.~\ref{fig18}.\\
Inserting eq.~\ref{eq_app_a6} into
eq.~\ref{eq_app_a4} and making use of
the relationship between $r_{0}$ and $\rho_{0}$ (eq.~\ref{eq_app_a2})
to eliminate $\rho_{0}$, one obtains:
\be
\left(\frac{\sigma^{2}}{G}\right)^{3/2}\frac{G^{1/2}r_{0}}{\sigma}\lambda
= cr_{0}^{1/\beta}
\ee
Finally, one arrives at an expression for $r_{0}$:
\be
r_{0} = \left(\frac{k_{B}\lambda}{Gm_{p}c}\right)^{\frac{\beta}{1-\beta}}T_{cl}^{\frac{\beta}{1-\beta}}
 \label{eq_app_a7}
\ee
For the values of $\beta = 0.28\pm 0.04$ and using for $c$ and
$\lambda$ the values given above, one obtains:
\be
r_{0} = 0.248\, T_{cl}^{0.3\bar{2}}\,\, pc
 \label{eq_app_a8}
\ee
and:
\be
\rho_{0} = 2.48\times 10^{18} \, {\rm M}_{\sun} {\rm Mpc}^{-3}
 \label{eq_app_a9}
\ee

\section[]{Cooling function} \label{append_2}
We adopt the cooling function provided by
\citet{1993ApJS...88..253S}, for a metallicity ${\rm [Fe/H] = -1}$,
which should be valid in the temperature interval $10^{4} < T < 10^{8}
{\rm K}$. However, the shocked IGM plasma in our simulations often reaches
temperatures higher than $10^{8} {\rm K}$, so we need to extend the
cooling function to higher temperatures.\\
\noindent
The cooling function for a fully ionized plasma in the trans-relativistic
regime ($5\times 10^{7}<T<5\times 10^{9}{\rm K}$, see \citealt{2006ApJ...637..313W}) has been computed by
\citet{1982ApJ...258..335S}, and includes contributions from different
relevant processes like $e^{+}e^{-}$ annihilations, and the related
bremsstrahlung rates. Note that, in principle, cooling from $e^{+}e^{-}$
annihilations is more than 2 orders of magnitudes higher than $e^{-}p$
and $e^{-}e^{-}$ bremsstrahlung cooling rates. However, in a
relativistic thermal plasma the production of $e^{+}e^{-}$ pairs
starts to be significant at temperatures larger than the threshold for
$\gamma\gamma \rightarrow e^{+}e^{-}$, i.e. for $T \geq 5.93\times
10^{9} {\rm K}$.\\
There exist few fits to Svensson's relativistic cooling
function. \citet{1983MNRAS.204.1269S} provide a fit with 4 
coefficients, each computed in tabulated form for different values of
the temperature. For the more limited temperature interval relevant to
our work, two approximate formulas are available from the
literature. \citep[][eqs. A1-A4]{2001ApJ...549..100P} provide the
following fitting formulae: 
\begin{eqnarray}
   \frac{\Lambda^{PO}}{n_{e}n{i}} = \sigma_T c \alpha_f m_e c^2 n_i^2
   \Big[
   \big\{ \lambda_{br}(T_e)
          + 6.0\times10^{-22}\theta_e^{-1/2} \big\}^{-1} \nonumber\\
         + \left(\frac{\theta_e}{4.82\times10^{-6}}\right)^{-12}
   \Big]^{-1},
\end{eqnarray}
where $\sigma_T$ is the Thomson cross section,
$\alpha_f$ the fine-structure constant, $n_i$ the number
density of ions, and $\theta_{e}\equiv kT_{e}/M_{e}c^{2}$. 
\begin{figure}
\centering
\includegraphics[scale=0.4]{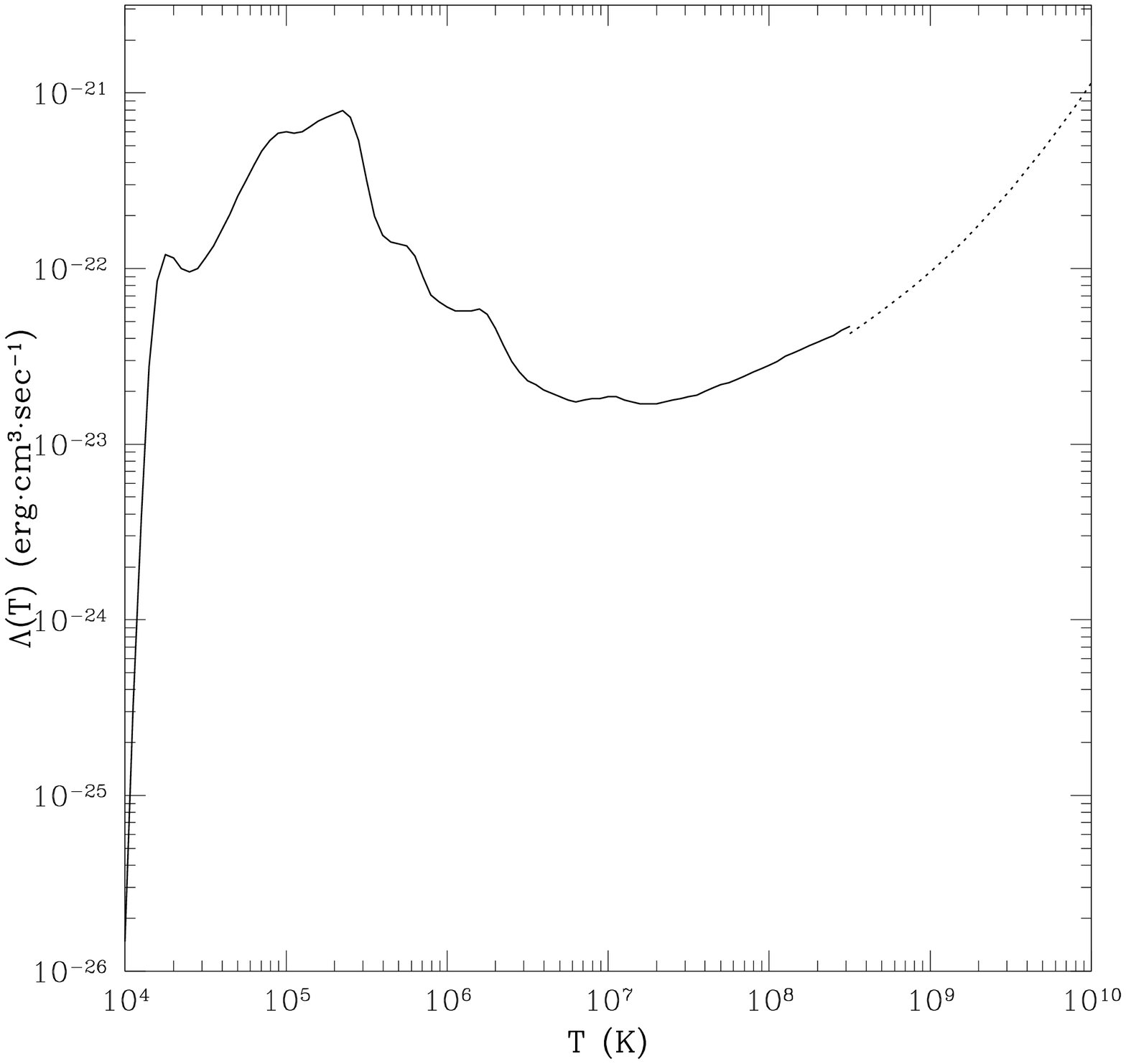}
\caption{The adopted cooling function. The continuous
  line denotes the Sutherland \& Dopita 1993 cooling function, while
  the dotted line is the approximation we have adopted for $T \geq
  10^{8.5}\,$ K, from Krause \& Alexander 2005.}
\label{fig19}
\end{figure}
The
bremsstrahlung cooling rate $\lambda_{br}(T_{e})$ is given by:
\begin{equation}
   \lambda_{br} =  \left(\frac{n_e}{n_i}\right)(\sum_i Z_i^2)
                   F_{ei}(\theta_e)
                 + \left(\frac{n_e}{n_i}\right)^2 F_{ee}(\theta_e)
\end{equation}
where
\begin{eqnarray*}
   F_{ei} & = & 4 \left(\frac{2}{\pi^3}\right)^{1/2}
                \theta_e^{1/2}(1+1.781\theta_e^{1.34})
          \quad\,\, \theta_e < 1 \\
          & = & \frac{9}{2\pi} \theta_e
                \left[\ln(1.123\theta_e+0.48)+1.5\right]
          \quad\,\, \theta_e > 1 \nonumber\\
   F_{ee} & = & \frac{5}{6\pi^{3/2}}(44-3\pi^2)\theta_e^{3/2}
                (1+1.1\theta_e+\theta_e^2-1.25\theta_e^{5/2})
          \quad\,\, \theta_e < 1 \\
          & = & \frac{9}{\pi} \theta_e
                \left[\ln(1.123\theta_e)+1.2746\right]
          \quad\,\, \theta_e > 1 .\nonumber
\end{eqnarray*}
This fit is claimed to be accurate to 2\% over the full
transrelativistic domain.\\
\noindent
A relatively simpler fit has been provided by \citet{2007MNRAS.376..465K}:
\be
\frac{\Lambda^{KA}}{n_{e}n_{i}} = 2.05\times 10^{-27}\sqrt{T}\left( 1 +
4.4\times10^{-10}T\right)
\ee
(Note that we have adopted units of $erg\cdot cm^{3}\cdot sec^{-1}$ in
both eq.s. 1 and 2). We show these fits in
fig.~\ref{fig19}. This latter 
figure should be compared with Fig. 9 of
\citet{1982ApJ...258..335S}. Svensson label the different curves
according to the {\em thickness parameter} $\tau_{n_{i}} =
n_{i}r_{e}^{2}R = 2.436\times 10^{-7} n_{i}\left[{\rm cm}^{-3}\right]
R\left[{\rm pc}\right]$, where: $r_{e}$ is the electron radius and $R$
is a typical size of the jet. It is clear that, for all the cases we
consider, one has $\tau_{n{i}} \ll 1$, so we can neglect the 
contribution from thermally-produced pairs to the cooling function,
which is also dominated by bremsstrahlung   in this temperature
range.\\
\noindent
Note however that no equilibrium relativistic plasma configuration is
possible for $\theta_{e} > 1$ (the curve labeled as ``O'' in Fig. 9
of \citet{1982ApJ...258..335S}, corresponding to $\theta_{max}=25$, or:
$T_{max}=1.48\times 10^{11} K$). At this critical maximum temperature
essentially all the plasma is converted to pairs, which then can leave
the volume and annihilate outside. For $T\geq T_{max}$ we then assume
$\lambda\rightarrow\infty, n_{e}\rightarrow 0$. The plasma is left
with ions, at a significantly lower temperature.


\label{lastpage}
\end{document}